\documentclass[runningheads]{llncs}

\usepackage[T1]{fontenc}
\usepackage{graphicx}
\usepackage{amsmath}
\usepackage{amssymb}
\usepackage{booktabs}
\usepackage{multirow}
\usepackage{array}
\usepackage{threeparttable}
\usepackage[table]{xcolor}
\usepackage{float}

\begin{document}

\title{Breaking the Quality--Intelligibility Trade-off in Streaming Target Speaker Extraction via Deep-Feature-Anchored Preference Optimization}
\titlerunning{Deep-Feature-Anchored DPO for Streaming TSE}
\author{Shuhai Peng\inst{1}\thanks{Equal contribution.},
Jinjiang Liu\inst{1}\protect\footnotemark[1],
Hui Lu\inst{2},
Liyang Chen\inst{1},
Guiping Zhong\inst{3},
\\
Jiakui Li\inst{3},
Shiyin Kang\inst{3},
Zhiyong Wu\inst{1}\thanks{Corresponding author.}}

\makeatletter
\def\@fnsymbol#1{\ensuremath{\ifcase#1\or *\or \dagger\fi}}
\makeatother

\authorrunning{S. Peng et al.}

\institute{Tsinghua University, \and
The Chinese University of Hong Kong, \and
SenseTime }

\maketitle
\vspace{-2em}
\raggedbottom

\begin{abstract}

    Generative streaming models for Target Speaker Extraction (TSE) commonly exhibit a quality--intelligibility trade-off, wherein naive optimization for perceptual audio quality tends to degrade speech intelligibility, and conversely.
    We reveal that this trade-off arises not from the constraints of streaming architectures, but from an inappropriate choice of optimization anchor. Directly optimizing against audio quality metrics
    induces catastrophic reward hacking, where content critical to pronunciation and intelligibility
    is systematically erased to maximize a proxy score.
    To break this bottleneck, we propose two complementary improvements:
    an enlarged Conformer convolution kernel for richer local spectro-temporal modeling,
    and WavLM-anchored Direct Preference Optimization (DPO) fine-tuning
    strategy.
    DPO preference pairs are ranked by WavLM cosine similarity, a deep acoustic feature encoding both
    phonetic structure and speaker identity, providing an optimization anchor that resists hacking.
    Under a 560\,ms streaming chunk size, the proposed method achieves a 10.9\% relative
    intelligibility improvement (word error rate: 0.138\,$\rightarrow$\,0.123), with marginal
    simultaneous gains in audio quality and speaker similarity.

\end{abstract}

\section{Introduction}

Target Speaker Extraction (TSE) aims to isolate the speech of a designated speaker from a complex acoustic mixture of interfering speakers and background noise~\cite{zmolikova2023neural}. Unlike blind source separation, which treats all concurrent sources symmetrically, TSE leverages auxiliary reference cues, typically a short enrollment utterance of the target speaker, to focus on a single acoustic target. This capability underpins critical real-world applications including teleconferencing systems, voice-controlled assistants, and automatic speech recognition in multi-talker environments.

For decades, the TSE domain was dominated by discriminative approaches such as SpEx+~\cite{ge2020spex} and WeSep~\cite{wang2024wesep}, which estimate time-frequency masks or filters to suppress interference. While computationally efficient, these methods introduce processing artifacts and struggle to reconstruct missing spectral details. More recently, generative models have catalyzed a paradigm shift: by treating speech extraction as a conditional generation task, language-model-based generative architectures such as TSELM-L~\cite{tang2024tselm} and LauraTSE~\cite{tang2025lauratse} achieve substantially higher audio fidelity compared to discriminative counterparts. However, these generative models rely on global bidirectional context that is fundamentally incompatible with real-time streaming deployments, where only past and current information is available.

The streaming-aware StarTSE~\cite{peng2026startsestreamingtargetspeaker}
established that autoregressive generative backbones could be adapted for streaming scenarios through chunk-wise interleaved splicing. However, adapting these models to streaming TSE exposes a critical tension intrinsic to real-time generation: the \textbf{quality--intelligibility trade-off}. A generative policy tuned strictly for acoustic smoothness will suppress the high-frequency bursts of stop consonants and fricatives, yielding high perceptual scores but severe intelligibility collapse. Conversely, optimizing purely for WER degrades perceptual audio quality. In the streaming setting, word error rate (WER) serves as the primary deployment criterion for intelligibility: it reflects the intelligibility and accuracy with which the model extracts the target speaker's speech from the mixture speech.

Preference-based fine-tuning, especially Direct Preference Optimization (DPO) \cite{rafailov2024direct}, offers a principled mechanism to improve the performance of generative streaming models without reinforcement learning~\cite{sutton1998reinforcement} overhead. However, the effectiveness of DPO is critically sensitive to the choice of optimization anchor: an ill-chosen ranking criterion can trigger \textit{reward hacking}, where the model exploits proxy metric blind spots and degrades attributes it was not directly optimized for. Our central objective is therefore to identify the optimal preference ranking criterion that delivers effective WER improvement while simultaneously preserving or improving audio quality and speaker similarity. To this end, we compare three DPO variants with different preference ranking criteria: DPO$_{\text{DNSMOS}}$ (perceptual scores), DPO$_{\text{WER}}$ (transcription-based word error rates), and DPO$_{\text{WavLM}}$ (deep acoustic feature similarity). Among these, DPO$_{\text{WavLM}}$ successfully breaks the quality--intelligibility trade-off, achieving simultaneous gains in WER, audio quality, and speaker similarity.

We identify that breaking this trade-off requires decoupling semantic preference learning from acoustic reconstruction. Since WER is governed by the semantic pathway's discrete token output, restricting gradient updates to this pathway directly targets intelligibility while leaving acoustic reconstruction intact. The contributions are twofold:
\begin{enumerate}
    \item \textbf{Diagnosing Reward Hacking in Streaming TSE.} We reveal that directly
    optimizing surface-level perceptual metrics (DNSMOS) induces severe reward hacking:
    the model learns to produce smooth, noise-free audio by suppressing stop consonants
    and fricatives, which raises perceptual scores but destroys the phonetic content
    essential for intelligibility, causing significant WER degradation.

    \item \textbf{Two-Stage Improvement for Streaming TSE.} We first show that enlarging
    the Conformer convolution kernel to $k{=}15$ provides richer temporal
    context within the streaming scenario, yielding a standalone WER reduction.
    Building on this, we propose WavLM-anchored DPO, which ranks preference pairs by
    deep acoustic feature similarity to provide an optimization anchor that resists
    reward hacking. Under 560\,ms chunk size, the combined approach achieves a 10.9\%
    relative WER reduction (0.138\,$\rightarrow$\,0.123) with simultaneous gains in
    audio quality and speaker similarity.
\end{enumerate}

\section{Method}
\label{sec:method}

The proposed method builds on the StarTSE streaming backbone and improves it in two stages: first, enlarging the Conformer convolution kernel to provide richer local temporal context; second, applying WavLM-anchored DPO fine-tuning to break the quality--intelligibility trade-off.

\subsection{Model Architecture}

\begin{figure}[t]
\centering
\includegraphics[width=\textwidth]{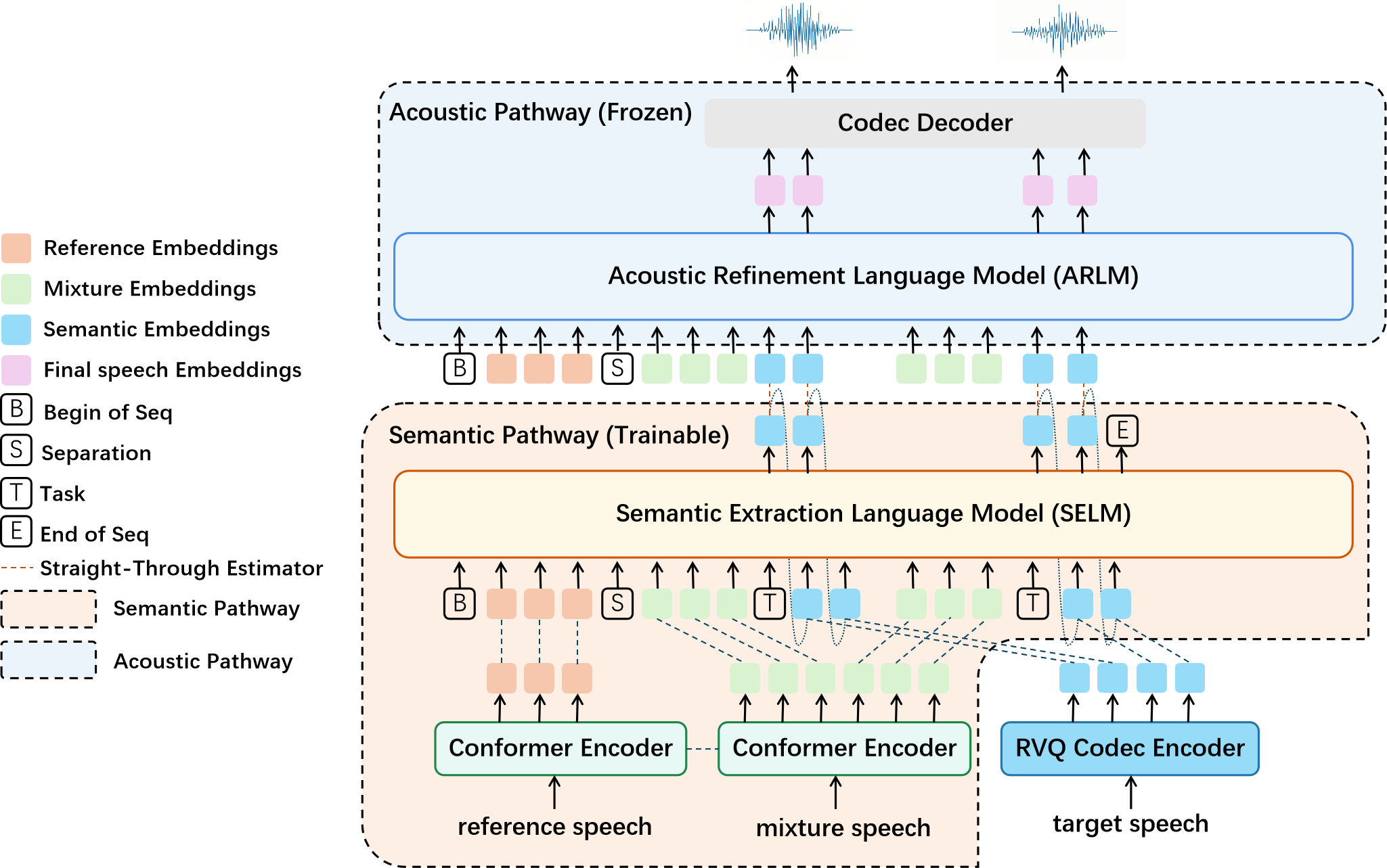}
\caption{Overview of the StarTSE architecture with the proposed DPO alignment strategy. The semantic pathway (Shared Conformer Encoder and SELM, highlighted in blue) receives gradient updates during DPO, while the acoustic pathway (ARLM and Codec Decoder, shaded in gray) remains frozen.}
\label{fig:architecture}
\end{figure}

The proposed framework adopts an encoder-decoder backbone operating under strict causal streaming constraints, as illustrated in Fig.~\ref{fig:architecture}. Architecturally, the model can be decomposed into two functionally distinct pathways: a \textbf{semantic pathway} responsible for preference-sensitive discrete token generation, and an \textbf{acoustic pathway} responsible for continuous-domain codec embedding reconstruction and waveform synthesis.

\paragraph{Semantic Pathway.}
This pathway comprises two components:
\begin{enumerate}
    \item \textbf{Shared Conformer Speech Encoder}: A causal Conformer encoder that processes both the input mixture Mel-spectrogram and the target speaker enrollment utterance through a shared streaming architecture. Unlike the Transformer encoder used in the original StarTSE, the Conformer introduces a convolution module within each block that captures local spectro-temporal patterns alongside the global self-attention mechanism. A key parameter is the convolution kernel size $k$, which governs the temporal receptive field of the local feature extraction.
    \item \textbf{Semantic Extraction Language Model (SELM)}: An autoregressive language model responsible for capturing semantic information from the mixture. To enable streaming, the SELM adopts a Chunk-wise Interleaved Splicing paradigm: the mixture embedding is segmented into chunks, and at each step $t$ the input is constructed by interleaving the static reference prefix with the historical sequence of mixture chunks and predicted semantic tokens. This enforces strict causality by limiting the receptive field to past observations. The SELM autoregressively generates discrete semantic tokens $U = \{u^{(1)}, \ldots, u^{(T)}\}$ chunk by chunk, which are subsequently passed to the acoustic pathway for waveform reconstruction.
\end{enumerate}

\paragraph{Acoustic Pathway.}
This pathway comprises two components:
\begin{enumerate}
    \item \textbf{Acoustic Refinement Language Model (ARLM)}: A language model that recovers fine-grained acoustic details from the SELM's output. Like the SELM, the ARLM operates on chunk-level interleaved inputs: at each step $t$, it interleaves mixture chunks with the corresponding SELM semantic embeddings $U_{\text{SELM}}^{(t)}$, appended to a static reference prefix. The ARLM outputs acoustically refined hidden representations $h^{(t)}$ that encode spectral and phase details complementary to the coarse semantic tokens. Critically, the SELM's logits are passed through a Straight-Through Estimator (STE) to produce differentiable soft embeddings, enabling end-to-end gradient flow from the acoustic reconstruction loss back into the semantic pathway.
    \item \textbf{Codec Decoder with Historical Context Refinement}: To address the boundary discontinuity inherent in chunk-wise streaming generation, the codec decoder explicitly leverages the output from the previous chunk. Instead of decoding each chunk in isolation, the decoder receives the concatenation of the current and previous ARLM hidden states $\text{Concat}(h^{(t-1)}, h^{(t)})$ as input. This Historical Context Refinement mechanism maintains continuous phase and semantic context across chunk boundaries, significantly enhancing speech quality under low-latency constraints.
\end{enumerate}

\subsection{Direct Preference Optimization}

Standard autoregressive training with token-level cross-entropy loss exhibits a well-known disconnect from human perceptual judgment: minimizing negative log-likelihood over discrete acoustic tokens does not guarantee natural-sounding speech. Direct Preference Optimization (DPO)~\cite{rafailov2024direct} addresses this gap by reformulating preference alignment as a contrastive classification problem, bypassing the need for explicit reward modeling or reinforcement learning.

Let $\pi_{\text{ref}}$ denote the pretrained reference policy and $\pi_\theta$ the policy being optimized. Since WER is determined by the discrete semantic tokens produced by the SELM, we restrict $\pi_\theta$ to the semantic pathway (Shared Encoder and SELM) and keep the acoustic pathway (ARLM and Codec Decoder) frozen at its pre-trained state. Given an input mixture $x$, a pair of generated outputs $(y_w, y_l)$ is constructed where $y_w$ is preferred over $y_l$ according to a specified criterion. The DPO objective is:

\begin{equation}
\mathcal{L}_{\text{DPO}}(\pi_\theta; \pi_{\text{ref}}) = -\mathbb{E}_{(x, y_w, y_l)} \left[ \log \sigma \left( \beta \log \frac{\pi_\theta(y_w|x)}{\pi_{\text{ref}}(y_w|x)} - \beta \log \frac{\pi_\theta(y_l|x)}{\pi_{\text{ref}}(y_l|x)} \right) \right]
\label{eq:dpo}
\end{equation}

where $\sigma(\cdot)$ is the logistic sigmoid and $\beta$ controls the KL-divergence penalty from $\pi_{\text{ref}}$. A larger $\beta$ constrains $\pi_\theta$ to remain closer to the pretrained distribution; a smaller $\beta$ allows more aggressive optimization at the risk of semantic drift.

\paragraph{Preference Pair Construction.}
For each mixture $x$, the pre-trained StarTSE model generates 16 candidate outputs by running the streaming inference procedure multiple times with stochastic sampling-based decoding. Each candidate is evaluated along three dimensions:
\begin{enumerate}
    \item \textbf{Perceptual Quality}: DNSMOS~\cite{reddy2021dnsmos} OVRL.
    \item \textbf{Intelligibility}: WER via Whisper LargeV3~\cite{radford2023robust}.
    \item \textbf{Speaker Similarity}: cosine similarity of WavLM~\cite{chen2022wavlm} embeddings against a clean reference from the target speaker.
\end{enumerate}

A central insight of this work is that the choice of metric used to rank preference pairs critically determines whether DPO converges to a Pareto-optimal~\cite{ehrgott2005multicriteria} solution or collapses via reward hacking. To isolate this effect, we generate a single set of candidate outputs from the same mixture inputs, and construct three independent DPO preference pair sets by ranking these identical candidates according to different criteria, as illustrated in Fig.~\ref{fig:dpo_ranking}. For each variant, the top-ranked candidate under its criterion is selected as $y_w$ and the bottom-ranked as $y_l$, forming a preference pair $(y_w, y_l)$:

\begin{figure}[ht]
\centering
\includegraphics[width=\linewidth]{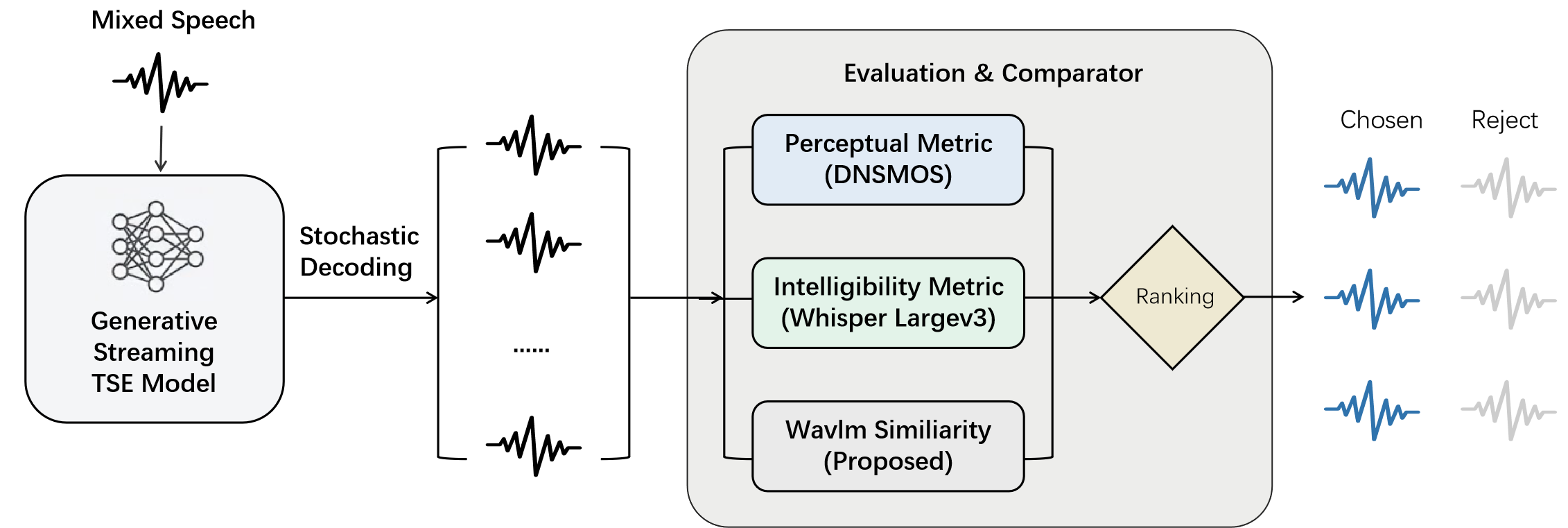}
\caption{Preference pair construction pipeline. For each mixture, 16 candidates are generated via stochastic decoding and evaluated along three dimensions (DNSMOS, WER, WavLM). The candidates are ranked independently by each of the three criteria; the top-ranked candidate is selected as $y_w$ and the bottom-ranked as $y_l$, producing three distinct preference pairs from the same underlying outputs.}
\label{fig:dpo_ranking}
\end{figure}

\paragraph{DPO$_{\text{DNSMOS}}$ (Perceptual-Driven).}
Pairs are ranked by DNSMOS OVRL, directly optimizing for perceptual naturalness. This variant tests whether quality scores alone provide a viable alignment signal.

\paragraph{DPO$_{\text{WER}}$ (Intelligibility-Driven).}
Pairs are ranked by WER (lower WER ranks higher), anchoring optimization on transcription accuracy. This variant prioritizes semantic preservation over perceptual enhancement.

\paragraph{DPO$_{\text{WavLM}}$ (Proposed, Deep-Feature-Driven).}
Pairs are ranked by WavLM cosine similarity against a clean reference. Deep self-supervised representations natively encode phonetic structure and speaker identity, providing an implicit semantic regularizer that DNSMOS metrics cannot replicate.

All variants use the standard DPO loss (Eq.~\ref{eq:dpo}) and share an identical training recipe differing only in the DPO metric. Each variant is evaluated across five KL penalty coefficients $\beta \in \{0.1, 0.3, 0.5, 0.7, 0.9\}$ and three checkpoints ($e_0$, $e_1$, $e_2$); the complete sweep across all $\beta$--checkpoint combinations is reported in Table~\ref{tab:full_results} of the Appendix. All DPO runs use the StarTSE model with $k{=}15$ Conformer convolution kernel as the starting checkpoint.

\subsection{Streaming Inference}

At inference time, the DPO-aligned model follows the same streaming protocol as the baseline StarTSE~\cite{peng2026startsestreamingtargetspeaker}: autoregressive token generation under 560\,ms latency with chunk-wise processing, followed by streaming vocoding. The proposed DPO-aligned model (with $k{=}15$ Conformer kernel) achieves a real-time factor (RTF) of 0.181 on an NVIDIA L40S GPU and 0.373 on an NVIDIA V100 GPU, both well below the 1.0 threshold for real-time streaming deployment.

\section{Experimental Setup}

\subsection{Dataset}

Experiments are conducted on the Libri2Mix corpus~\cite{cosentino2020librimix}, a two-speaker mixture dataset derived from LibriSpeech 460h~\cite{panayotov2015librispeech}. All audio is resampled to 16\,kHz mono. The training, validation, and test sets contain 64,700, 3,000, and 3,000 utterances, respectively. DPO preference pairs are pre-generated offline following the procedure described in Section~\ref{sec:method}.

\subsection{Implementation Details}

The model architecture follows StarTSE but replaces the original Transformer encoder with a 6-block causal Conformer encoder (hidden dimension 512, 8 attention heads, convolution kernel $k{=}15$) to capture richer local spectro-temporal patterns. The remaining components are unchanged: a 10-layer decoder-only SELM transformer and a 6-block ARLM Conformer. The FunCodec neural codec~\cite{du2023funcodec} is used with 32 quantizers and a codebook size of 1,024 at 25\,Hz token rate.

DPO training uses the Adam optimizer with a learning rate of $1 \times 10^{-5}$ warmed up over 200 steps. Five KL penalty coefficients $\beta \in \{0.1, 0.3, 0.5, 0.7, 0.9\}$ and three checkpoints ($e_0$, $e_1$, $e_2$) are evaluated per DPO variant. All experiments run on 10 nodes, each equipped with 8 NVIDIA V100-32GB GPUs.

\subsection{Evaluation Metrics}

We evaluate models along four dimensions:
\begin{itemize}
    \item \textbf{Perceptual Quality}: DNSMOS P.835~\cite{reddy2021dnsmos} scores---SIG (signal quality), BAK (background quality), and OVL (overall quality).
    \item \textbf{Intelligibility}: Word Error Rate (WER) computed via Whisper LargeV3~\cite{radford2023robust}.
    \item \textbf{Speaker Similarity}: Cosine similarity of WavLM~\cite{chen2022wavlm} and WeSpeaker~\cite{wang2022wespeaker} embeddings against a clean reference.
    \item \textbf{Inference Success Rate (ISR)}: The percentage of test samples for which autoregressive generation completes without collapse, defined as
    \begin{equation}
    \text{ISR} = \frac{1}{N} \sum_{i=1}^{N} \mathbb{I}(\text{valid}(\hat{y}_i)) \times 100\%,
    \end{equation}
    where $\mathbb{I}(\cdot)$ is the indicator function and $\text{valid}(\hat{y}_i)$ is true if generation succeeds.
\end{itemize}

\subsection{Baselines}

The primary baseline is StarTSE~\cite{peng2026startsestreamingtargetspeaker}, a streaming-aware generative TSE model. For all DPO experiments, the reference policy $\pi_{\text{ref}}$ is StarTSE with the Conformer encoder kernel enlarged to $k{=}15$ (the stage-1 improvement described in Section~\ref{sec:method}). Discriminative methods (SpEx+~\cite{ge2020spex}, WeSep~\cite{wang2024wesep}) and generative methods (TSELM-L~\cite{tang2024tselm}, LauraTSE~\cite{tang2025lauratse}) are included in the overall comparison.

\section{Experimental Results and Discussion}

\subsection{Main Results}
\label{sec:overall}

Table~\ref{tab:main} reports results at a 560\,ms streaming chunk size. Methods are categorized by paradigm: discriminative (D) vs.\ generative (G), and by mode (Off. = offline, non-causal). StarTSE serves as the base model for all subsequent DPO experiments (Section~\ref{sec:dpo_main} and the ablation studies).

\begin{table}[ht]
\caption{Comparison with existing TSE methods. Offline methods serve as upper-bound references. D = discriminative; G = generative. Bold: best among streaming; underline: second-best among streaming.}
\label{tab:main}
\centering
\renewcommand{\arraystretch}{1.15}
\setlength{\tabcolsep}{1.8pt}
\footnotesize
\begin{tabular}{l|llrrrrrrr}
\hline
\multirow{2}{*}{\shortstack{Inference\\Mode}} & \multirow{2}{*}{Method} & \multirow{2}{*}{Cat.} &
  \multicolumn{3}{c}{DNSMOS$\uparrow$} &
  \multirow{2}{*}{WER$\downarrow$} &
  \multicolumn{2}{c}{Sim$\uparrow$} & \multirow{2}{*}{ISR$\uparrow$} \\
& & & SIG & BAK & OVL & & WavLM & WeSpk & \\
\hline
\multirow{5}{*}{\small Offline} &
  Mixture & -- & 3.383 & 3.098 & 2.653 & 0.580 & 0.847 & 0.759 & 100.00\% \\
& SpEx+~\cite{ge2020spex} & D & 3.472 & 4.027 & 3.186 & -- & 0.973 & 0.935 & -- \\
& WeSep~\cite{wang2024wesep} & D & 3.486 & 3.838 & 3.118 & -- & 0.980 & 0.945 & -- \\
& TSELM-L~\cite{tang2024tselm} & G & 3.489 & 4.041 & 3.212 & -- & 0.887 & 0.627 & -- \\
& LauraTSE~\cite{tang2025lauratse} & G & 3.607 & 4.078 & 3.336 & 0.082 & 0.973 & 0.874 & 100.00\% \\
\hline
\multirow{2}{*}{\small Streaming} &
  LauraTSE~\cite{tang2025lauratse} & G & 3.477 & \textbf{3.879} & \textbf{3.130} & 0.174 & 0.954 & 0.739 & 99.10\% \\
& StarTSE~\cite{peng2026startsestreamingtargetspeaker} & G & \textbf{3.535} & 3.752 & 3.117 & \underline{0.138} & \underline{0.959} & \underline{0.847} & \textbf{100.00\%} \\
\rowcolor{gray!20} & \textbf{Proposed} & G & \underline{3.491} & \underline{3.837} & \underline{3.122} & \textbf{0.123} & \textbf{0.965} & \textbf{0.853} & \textbf{100.00\%} \\
\hline
\end{tabular}
\end{table}

Several observations emerge from Table~\ref{tab:main}. The offline setting establishes upper bounds: LauraTSE achieves WER of 0.082 and OVL of 3.336 with full bidirectional context, yet under the 560\,ms streaming constraint its WER degrades sharply to 0.174 and ISR drops to 99.10\%, illustrating the inevitable degradation when a non-streaming-aware model is forced into causal inference. StarTSE~\cite{peng2026startsestreamingtargetspeaker} closes much of this gap: at 560\,ms, it achieves WER of 0.138---a 20.7\% relative reduction over streaming LauraTSE---along with substantially higher speaker similarity (WavLM 0.959 vs.\ 0.954, WeSpk 0.847 vs.\ 0.739), and maintains 100\% ISR. Notably, StarTSE matches the offline WeSep in OVL (3.117 vs.\ 3.118) while operating under strict causal constraints.

The proposed DPO-aligned model (bottom row, DPO$_{\text{WavLM}}$) further improves WER to \textbf{0.123}, a 10.9\% relative gain over the StarTSE baseline and a 29.3\% reduction relative to LauraTSE (0.174), with simultaneous improvements in BAK (3.752$\rightarrow$3.837) and both speaker similarity metrics.

\subsection{DPO Alignment: Breaking the Quality--Intelligibility Bottleneck}
\label{sec:dpo_main}

Table~\ref{tab:best_results} reports the results of StarTSE with the $k{=}15$ Conformer kernel under different DPO metrics:

The proposed DPO$_{\text{WavLM}}$ breaks the quality--intelligibility trade-off. At $\beta=0.9$, $e_0$, WER reaches \textbf{0.123}---a 10.9\% relative reduction over the bare baseline (0.138). This gain splits roughly evenly: the $k=15$ Conformer kernel contributes 6.5\% (0.138$\to$0.129), and DPO$_{\text{WavLM}}$ adds 4.7\% (0.129$\to$0.123). Critically, perceptual quality improves \textit{simultaneously}: DNSMOS OVL rises from 3.117 to 3.122, BAK from 3.752 to 3.837, and WavLM similarity from 0.959 to 0.965. Thus, DPO anchored by the WavLM criterion achieves a Pareto improvement over the baseline: the optimization simultaneously advances WER, perceptual quality, and speaker similarity, without sacrificing any dimension.

DPO$_{\text{DNSMOS}}$ achieves the highest perceptual scores (OVL 3.257, SIG 3.606, BAK 3.908) but at devastating semantic cost: WER degrades to 0.161, erasing the kernel enlargement gain. 

DPO$_{\text{WER}}$ yields no intelligibility gain over the $k{=}15$ baseline (0.130 vs.\ 0.129) and degrades perceptually (OVL 3.110 vs.\ 3.121). Only DPO$_{\text{WavLM}}$ achieves both objectives, ranking either best or second-best across every metric.

\begin{table}[ht]
\centering
\caption{Overall performance of StarTSE ($k=15$ Conformer kernel) under different DPO DPO metrics. Bold: best; underline: second-best.}
\label{tab:best_results}
\footnotesize
\begin{tabular}{l rrr r rr}
\toprule
\multirow{2}{*}{Strategy} &
  \multicolumn{3}{c}{DNSMOS$\uparrow$} & \multirow{2}{*}{WER$\downarrow$} & \multicolumn{2}{c}{Sim$\uparrow$} \\
& SIG & BAK & OVL & & WavLM & WeSpk \\
\midrule
StarTSE & 3.535 & 3.752 & 3.117 & 0.138 & 0.959 & 0.847 \\
$+$ $k{=}15$ kernel & \underline{3.541} & 3.752 & 3.121 & \underline{0.129} & 0.961 & 0.848 \\
\quad $+$ DPO$_{\text{DNSMOS}}$ & \textbf{3.606} & \textbf{3.908} & \textbf{3.257} & 0.169 & 0.953 & 0.823 \\
\quad $+$ DPO$_{\text{WER}}$ & 3.521 & 3.763 & 3.110 & 0.130 & \textbf{0.966} & 0.855 \\
\rowcolor{gray!20} \quad $+$ DPO$_{\text{WavLM}}$ & 3.491 & \underline{3.837} & \underline{3.122} & \textbf{0.123} & \underline{0.965} & 0.853 \\
\bottomrule
\end{tabular}
\end{table}

\begin{figure}[t]
\centering
\begin{minipage}[b]{0.49\linewidth}
  \centering
  \includegraphics[width=\linewidth]{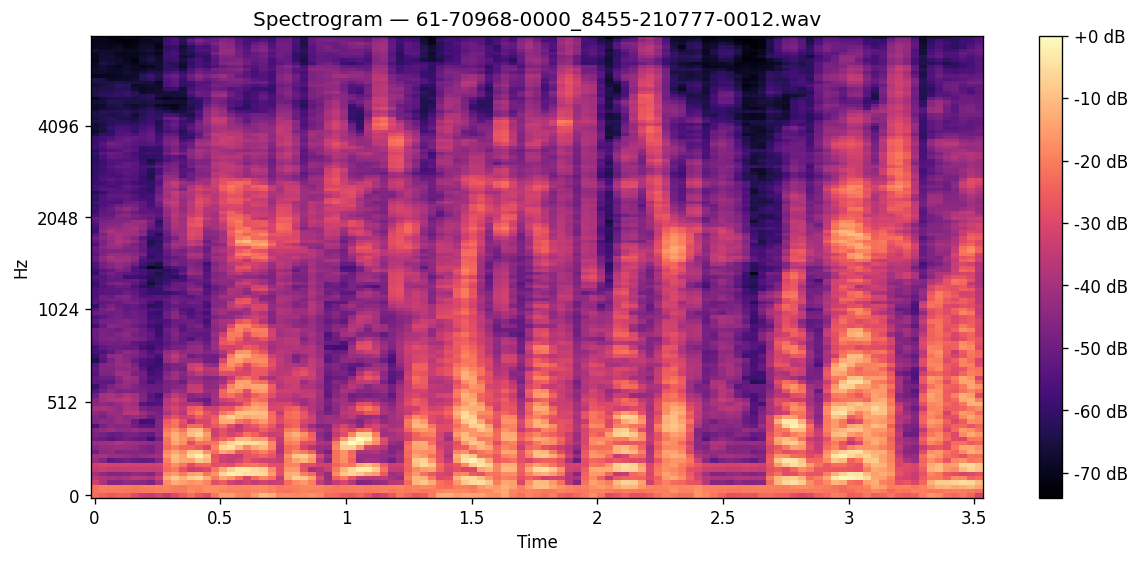}
  \vspace{0.3ex}
  \small (a) Ground Truth
\end{minipage}
\hfill
\begin{minipage}[b]{0.49\linewidth}
  \centering
  \includegraphics[width=\linewidth]{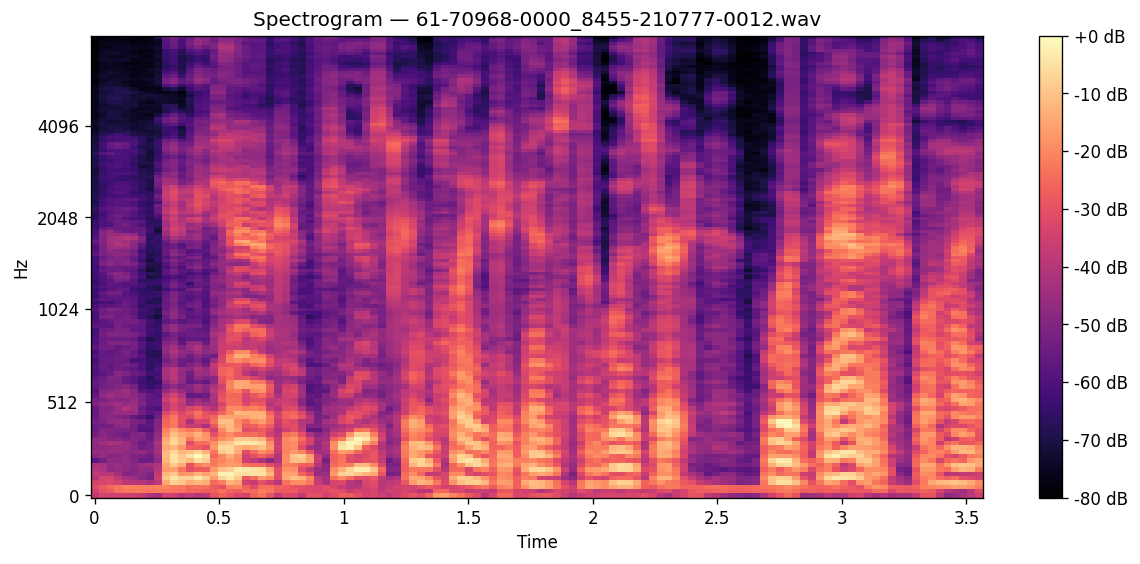}
  \vspace{0.3ex}
  \small (b) DPO$_{\text{WavLM}}$ (Proposed)
\end{minipage}
\vspace{1ex}

\begin{minipage}[b]{0.65\linewidth}
  \centering
  \includegraphics[width=\linewidth]{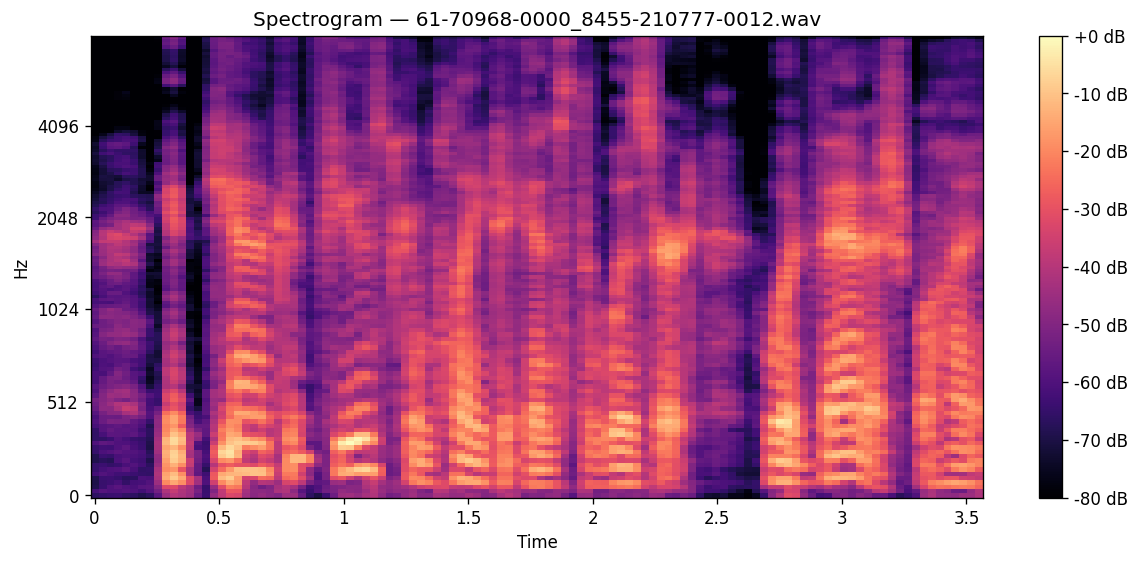}
  \vspace{0.3ex}
  \small (c) DPO$_{\text{DNSMOS}}$ --- reward hacking: low-frequency attenuation and erased transients
\end{minipage}
\caption{Mel spectrograms of the same utterance under three conditions (same sample as Table~\ref{tab:best_results}).
DPO$_{\text{WavLM}}$ (b) closely matches the ground truth (a), preserving
low-frequency fundamentals and the mid-frequency harmonic series.
DPO$_{\text{DNSMOS}}$ (c) shows severe low-frequency attenuation
(0--512\,Hz region darkened) and erased consonantal transients (black
voids at mid-to-high frequencies): the reward-hacking policy suppresses
acoustic content that the DNSMOS evaluator misidentifies as distortion,
yielding spectrally smooth but semantically vacuous speech---directly
explaining the WER collapse reported in Table~\ref{tab:best_results}.}
\label{fig:mel}
\end{figure}

\subsection{How Deep Features Break the Trade-off}
\label{sec:reward_hacking}

Table~\ref{tab:best_results} reports DPO performance under three different DPO metrics. Despite sharing an identical training recipe, the three variants diverge sharply. DPO$_{\text{DNSMOS}}$ produces the highest perceptual scores in the table yet suffers severe WER degradation. DPO$_{\text{WER}}$ yields no intelligibility improvement and degrades perceptual quality. Only DPO$_{\text{WavLM}}$ improves both dimensions simultaneously. The complete set of results across all $\beta \in \{0.1, 0.3, 0.5, 0.7, 0.9\}$ and checkpoints $e_0$, $e_1$, $e_2$ is provided in Table~\ref{tab:full_results} in the Appendix.

\paragraph{DPO$_{\text{DNSMOS}}$: intelligibility collapse despite high perceptual scores.}
The first row of DPO results in Table~\ref{tab:best_results} presents a contradiction: DNSMOS OVRL reaches 3.257, the highest perceptual score among all variants, yet WER degrades to 0.169 at $e_0$ and continues to deteriorate with training---reaching 0.316 by $e_2$ at $\beta{=}0.9$, and exceeding 0.8 at lower $\beta$. Figure~\ref{fig:mel} reveals the physical cause: DPO$_{\text{DNSMOS}}$ (c) suppresses low-frequency fundamentals (0--512\,Hz) and erases consonantal transients relative to the ground truth (a). DNSMOS evaluates only waveform smoothness and is blind to whether linguistic content is preserved. The gradient therefore actively penalizes the very acoustic features---stop bursts, plosives, fricatives---that carry phonetic information. Optimizing against a proxy that encodes only waveform smoothness, with no notion of linguistic content, is therefore intrinsically one-sided reward hacking: the model maximizes the score by destroying the very phonetic structure that intelligibility requires.

\paragraph{DPO$_{\text{WER}}$: no intelligibility gain, lower quality.}
DPO$_{\text{WER}}$ ranks candidates solely by word error rate. The result is a net regression relative to the $k{=}15$ kernel alone (WER 0.129, OVL 3.121): even the best DPO$_{\text{WER}}$ configuration ($\beta{=}0.9$, $e_2$) reaches only WER 0.131, and OVL drops to 3.110. In other words, optimizing against a single transcription-based metric not only fails to improve intelligibility beyond what the enlarged kernel already provides, but also degrades perceptual quality. The WER signal anchors the encoder conservatively, preventing the representation from drifting, but provides no guidance on what constitutes natural speech. The optimization inadvertently shifts the representation away from the acoustic properties that produce natural-sounding output, revealing that a transcription-only metric is too narrow to direct the encoder toward regions that benefit both dimensions simultaneously.

\paragraph{DPO$_{\text{WavLM}}$: simultaneous gains.}
DPO$_{\text{WavLM}}$ ranks candidates by cosine similarity to WavLM representations of a clean reference. Unlike DNSMOS, which encodes only waveform smoothness, or WER, which encodes only transcription accuracy, WavLM's internal representations jointly capture speaker identity, phonetic structure, and acoustic continuity---in effect, encoding what real speech sounds like, having been pre-trained on large-scale natural speech. This has a decisive consequence for optimization: to maximize WavLM similarity, the encoder must produce features that are simultaneously phonetically discriminative and acoustically natural. The gradient is thus constrained by a prior that neither DNSMOS-based nor WER-based optimization provides. The result, shown in the final row of Table~\ref{tab:best_results}, is a Pareto improvement over the $k{=}15$ baseline: WER drops from 0.129 to \textbf{0.123} (a 4.7\% relative reduction attributed purely to DPO), while BAK rises from 3.752 to 3.837, OVL from 3.121 to 3.122, WavLM similarity from 0.961 to 0.965, and WeSpk similarity from 0.848 to 0.853. WER stays below 0.165 across all configurations, matching DPO$_{\text{WER}}$ in stability while exceeding it in peak performance. 

Table~\ref{tab:full_results} reveals how different preference anchors dictate optimization stability across varying KL constraints $\beta$. For $DPO_{DNSMOS}$, relaxed constraints ($\beta \le 0.3$) trigger catastrophic reward hacking, severely degrading intelligibility. Conversely, $DPO_{WER}$ remains overly conservative, failing to surpass the pre-trained baseline's performance regardless of the $\beta$ setting. $DPO_{WavLM}$ uniquely avoids both extremes: it provides a reliable optimization direction where stronger regularization ($\beta = 0.9$) steadily improves WER to a global best of 0.123. By jointly encoding phonetic structure and acoustic fidelity, deep features like WavLM constrain the representation shift to a region where semantic precision and acoustic naturalness coexist. This demonstrates the precise mechanism by which deep-feature anchoring breaks the quality-intelligibility trade-off

\subsection{Conformer Kernel Size Ablation}
\label{sec:cnn}

Replacing the Transformer encoder with a Conformer introduces a convolution module in each block, whose kernel size $k$ controls the local temporal receptive field. Table~\ref{tab:ablation_kernel} compares two Conformer configurations against the Transformer baseline.

\begin{table}[t]
\centering
\renewcommand{\arraystretch}{1.3}
\caption{Ablation on the Conformer convolution kernel size. Bold: best.}
\label{tab:ablation_kernel}
\footnotesize
\begin{tabular}{l rrr r rr}
\toprule
\multirow{2}{*}{Encoder} &
  \multicolumn{3}{c}{DNSMOS$\uparrow$} & \multirow{2}{*}{WER$\downarrow$} & \multicolumn{2}{c}{Sim$\uparrow$} \\
& SIG & BAK & OVL & & WavLM & WeSpk \\
\midrule
Transformer (StarTSE)  & 3.535 & 3.752 & 3.117 & 0.138 & 0.959 & 0.847 \\
Conformer ($k=3$)  & 3.521 & 3.744 & 3.101 & 0.152 & 0.955 & 0.837 \\
\rowcolor{gray!20} Conformer ($k=15$) & \textbf{3.541} & \textbf{3.752} & \textbf{3.121} & \textbf{0.129} & \textbf{0.961} & \textbf{0.848} \\
\bottomrule
\end{tabular}
\end{table}

The Transformer baseline (StarTSE) achieves WER of 0.138. Replacing it with a Conformer at $k{=}3$ degrades WER to 0.152 and OVL to 3.101---worse than the original Transformer across every metric. A kernel this small captures too little local context to complement the self-attention mechanism; the convolution module introduces noise rather than useful temporal structure. Increasing the kernel to $k{=}15$ reverses this: the larger receptive field spans complete syllabic structures---onset transitions, vowel nuclei, and coda resolutions---within a single convolution, allowing the convolution and self-attention pathways to become complementary. The result is WER of \textbf{0.129} (6.5\% reduction over the Transformer baseline) with simultaneous improvements in OVL (3.121), SIG (3.541), and both speaker similarity metrics.

%
%
%

\section{Conclusion}
Generative streaming models for target speaker extraction inherently face a quality-intelligibility trade-off. 
This paper demonstrates that this tension is not a fundamental limitation of streaming architectures, but rather 
a consequence of using inappropriate optimization anchors. To resolve this, we propose two complementary improvements: 
expanding the Conformer convolution kernel to $k=15$ to capture richer local context, and introducing a deep-feature-anchored 
Direct Preference Optimization (DPO) strategy. 

Crucially, we reveal that ranking DPO preference pairs via WavLM cosine similarity—a metric that jointly encodes phonetic structure and acoustic fidelity—prevents the reward hacking induced by purely perceptual metrics like DNSMOS and the acoustic degradation caused by transcription-only metrics like WER. This deep-feature anchoring successfully breaks the bottleneck, yielding a 10.9\% relative 
reduction in word error rate while simultaneously improving perceptual quality and speaker similarity.

Looking ahead, this research opens several promising avenues for future investigation. First, extending this deep-feature-anchored paradigm 
to lower-latency regimes will be critical for strict real-time deployments(<200ms). 
Second, evaluating and enhancing the robustness of this approach across more diverse acoustic environments and complex multi-speaker scenarios represents 
a necessary next step. 
Finally, an important open question is the development of multi-objective DPO strategies. Future frameworks could explicitly and 
dynamically balance perceptual quality, intelligibility, and speaker fidelity, effectively moving beyond the reliance on a single deep-feature anchor.

\appendix
\section{Full Experimental Results}
\begin{table}[H]
\centering
\caption{Complete DPO experimental results. All models use 560\,ms streaming latency.}
\label{tab:full_results}
\resizebox{\textwidth}{!}{
\begin{tabular}{l r r r r r r r}
\toprule
Strategy & $\beta$ & Ep. & DNSMOS SIG & DNSMOS BAK & DNSMOS OVL & WER & WavLM Sim \\
\midrule
StarTSE & -- & -- & 3.5350 & 3.7520 & 3.1170 & 0.1380 & 0.9590 \\
\midrule
\multirow{5}{*}{\textit{DNSMOS-first}} & 0.1 & e0 & 3.5814 & 3.8201 & 3.1911 & 0.1609 & 0.9547 \\
 & 0.1 & e1 & 3.3497 & 3.5239 & 2.9168 & 0.3866 & 0.9154 \\
 & 0.1 & e2 & 3.1381 & 3.2731 & 2.6878 & 0.5258 & 0.8895 \\
\cmidrule{2-8}
 & 0.3 & e0 & 3.3332 & 3.4532 & 2.9184 & 0.3809 & 0.9073 \\
 & 0.3 & e1 & 2.1570 & 1.7955 & 1.6898 & 0.7204 & 0.7921 \\
 & 0.3 & e2 & 1.8247 & 1.4737 & 1.4232 & 0.8126 & 0.7713 \\
\cmidrule{2-8}
 & 0.5 & e0 & 3.5985 & 3.8705 & 3.2315 & 0.1845 & 0.9475 \\
 & 0.5 & e1 & 2.5727 & 2.4188 & 2.1500 & 0.5966 & 0.8231 \\
 & 0.5 & e2 & 2.0907 & 1.7462 & 1.6484 & 0.7373 & 0.7706 \\
\cmidrule{2-8}
 & 0.7 & e0 & 3.6075 & 3.8630 & 3.2354 & 0.1683 & 0.9494 \\
 & 0.7 & e1 & 3.4980 & 3.7067 & 3.0947 & 0.4235 & 0.9102 \\
 & 0.7 & e2 & 3.4411 & 3.6889 & 3.0506 & 0.5204 & 0.8887 \\
\cmidrule{2-8}
 & 0.9 & e0 & \textbf{3.6063} & \textbf{3.9083} & \textbf{3.2567} & 0.1608 & 0.9525 \\
 & 0.9 & e1 & 3.6020 & 3.8859 & 3.2416 & 0.2237 & 0.9388 \\
 & 0.9 & e2 & 3.5825 & 3.8824 & 3.2223 & 0.3157 & 0.9244 \\
\midrule
\multirow{5}{*}{\textit{WER-first}} & 0.1 & e0 & 3.5260 & 3.7640 & 3.1140 & 0.1597 & 0.9603 \\
 & 0.1 & e1 & 3.5110 & 3.7540 & 3.1000 & 0.1378 & 0.9637 \\
 & 0.1 & e2 & 3.5090 & 3.7770 & 3.1080 & 0.1379 & 0.9637 \\
\cmidrule{2-8}
 & 0.3 & e0 & 3.5250 & 3.7870 & 3.1260 & 0.1490 & 0.9621 \\
 & 0.3 & e1 & 3.5080 & 3.7700 & 3.1030 & 0.1343 & 0.9649 \\
 & 0.3 & e2 & 3.5080 & 3.7640 & 3.0990 & 0.1369 & 0.9648 \\
\cmidrule{2-8}
 & 0.5 & e0 & 3.5270 & 3.7750 & 3.1210 & 0.1460 & 0.9631 \\
 & 0.5 & e1 & 3.5180 & 3.7630 & 3.1080 & 0.1359 & 0.9654 \\
 & 0.5 & e2 & 3.5100 & 3.7580 & 3.1000 & 0.1307 & 0.9657 \\
\cmidrule{2-8}
 & 0.7 & e0 & 3.5340 & 3.7930 & 3.1360 & 0.1423 & 0.9635 \\
 & 0.7 & e1 & 3.5220 & 3.7760 & 3.1170 & 0.1372 & 0.9655 \\
 & 0.7 & e2 & 3.5170 & 3.7600 & 3.1060 & 0.1329 & 0.9656 \\
\cmidrule{2-8}
 & 0.9 & e0 & 3.5310 & 3.7800 & 3.1260 & 0.1374 & 0.9639 \\
 & 0.9 & e1 & 3.5210 & 3.7720 & 3.1140 & 0.1343 & 0.9652 \\
 & 0.9 & e2 & 3.5210 & 3.7630 & 3.1100 & \textbf{0.1309} & \textbf{0.9660} \\
\midrule
\multirow{5}{*}{\textit{WavLM-first} (ours)} & 0.1 & e0 & 3.3913 & 3.8216 & 3.0321 & 0.1551 & 0.9593 \\
 & 0.1 & e1 & 3.4027 & 3.8049 & 3.0359 & 0.1600 & 0.9600 \\
 & 0.1 & e2 & 3.4010 & 3.7987 & 3.0309 & 0.1646 & 0.9597 \\
\cmidrule{2-8}
 & 0.3 & e0 & 3.4598 & 3.8356 & 3.0949 & 0.1378 & 0.9629 \\
 & 0.3 & e1 & 3.4470 & 3.8420 & 3.0889 & 0.1432 & 0.9632 \\
 & 0.3 & e2 & 3.4294 & 3.8527 & 3.0783 & 0.1475 & 0.9624 \\
\cmidrule{2-8}
 & 0.5 & e0 & 3.4778 & 3.8541 & 3.1184 & 0.1344 & 0.9643 \\
 & 0.5 & e1 & 3.4732 & 3.8542 & 3.1161 & 0.1371 & 0.9644 \\
 & 0.5 & e2 & 3.4640 & 3.8586 & 3.1098 & 0.1423 & 0.9641 \\
\cmidrule{2-8}
 & 0.7 & e0 & 3.4912 & 3.8511 & 3.1287 & 0.1290 & 0.9643 \\
 & 0.7 & e1 & 3.4887 & 3.8554 & 3.1293 & 0.1388 & 0.9646 \\
 & 0.7 & e2 & 3.4749 & 3.8602 & 3.1200 & 0.1472 & 0.9641 \\
\cmidrule{2-8}
 & 0.9 & e0 & 3.4909 & 3.8368 & 3.1220 & \textbf{0.1232} & 0.9652 \\
 & 0.9 & e1 & 3.4876 & 3.8436 & 3.1219 & 0.1293 & 0.9655 \\
 & 0.9 & e2 & 3.4790 & 3.8420 & 3.1142 & 0.1381 & 0.9651 \\
\bottomrule
\end{tabular}
}
\end{table}

\end{document}